# Secret Key Generation from Channel Noise with the Help of a Common Key

Tatsuya TOMARU[†]

**SUMMARY** Information-theoretically secure communications are possible when channel noise is usable and when the channel has an intrinsic characteristic that a legitimate receiver (Bob) can use the noise more advantageously than an eavesdropper (Eve). This report deals with the case in which the channel does not have such an intrinsic characteristic. Here, we use a pre-shared common key as a tool that extrinsically makes Bob more advantageous than Eve. This method uses error-correcting code in addition to the common key and noise, and manages the three components in random-number transmission. Secret keys are generated from noise, and messages are encrypted with the secret keys in a one-time pad manner. As a result, information leaks meaningful to Eve are restricted to the parity-check symbols for the random numbers. It is possible to derive the candidates of the common key from the parity check symbols, and the security of this method is quantified in terms of the amount of computations needed for an exhaustive search of the candidates, where we evaluate the security by assuming that all parity check symbols leak to Eve without bit errors. Noise contributes to not only generating secret keys but also enhancing the security because the candidates of the common key increase with it.
*key words:* Channel, noise, error-correcting code, common key, secret key, bit error

## 1. Introduction

Highly confidential information, e.g., government and military secrets, must be communicated with maximum security between a limited number of parties. This kind of information might affect national fortunes 50 or even 100 years hence, and therefore, its security must be long term. This report proposes a method that meets this requirement. It discusses security under the following three conditions: (1) only technologies available at present can be used; (2) communication is world-wide; (3) only a limited number of parties communicate with each other.

The method of Wyner [1] using channel noise is one way of maximizing security, and it achieves information-theoretic security. When the bit-error rate (BER) of an eavesdropper (Eve) is higher than that of a legitimate receiver (Bob), this difference generates a secrecy capacity [1,2]. Wyner assumed that Bob receives signals at a lower bit-error rate than Eve, but this assumption was later loosened. That is, so long as some of Eve's received errors are different from Bob's, Bob's receiving conditions don't have to be better than Eve's, and the amount of information corresponding to Eve's unique errors can be transformed into secret keys through public discussions between the sender (Alice) and Bob[1] [3,4]. The method using channel noise has since been widely studied, and the researches can classified into ones on channel-type models and ones on source-type models [4]. The former model is one in which Alice and Bob share randomness from the channel noise. It requires bit errors that only Eve suffers from in order to generate secret keys from the noise [5]. A broadcast channel is usually assumed to conform to this situation [2,4]. The latter model is one in which both Alice and Bob receive randomness from a source, and when the randomness is correlated between Alice and Bob, independently of Eve, secret keys are generated [6]. Approaches that take both models into consideration have also been studied [7–9].

The noisy channel method can achieve information-theoretic security. Another method that does so is quantum cryptography [10–12]. Quantum cryptography similarly possesses the concept of channel-type and source-type models. The BB84 protocol, wherein single photons are transmitted and received, corresponds to a channel-type model [10]. Quantum entanglement-related methods correspond to source-type models [13,14].

While methods having information-theoretic security are achievable in principle, they are difficult to apply to long-haul optical fiber transmissions. Optical fibers cause transmission losses of 0.2 dB/km, and as a result, quantum cryptography using single photons is limited to about 100 km. The noisy channel method faces another difficulty. Eavesdropping is easy in fiber communications if there is a preinstalled photo-coupler that divides the light. In particular, if the photo-coupler is near the transmitter, Eve can receive signals without being affected by channel noise. In this case, it cannot be assumed that Eve will have unique errors, and thus, the noisy channel model cannot be applied.

---

[†]The author is with Center for Exploratory Research, Research & Development Group, Hitachi, Ltd., Hatoyama, Saitama, 350-0395 Japan.
Email: tatsuya.tomaru.yq@hitachi.com

[1] As a result of public discussions, Bob gains an advantage over Eve.

Both the noisy channel method and quantum cryptography have an important mechanism wherein Bob can become more advantageous than Eve; the noisy channel method uses a broadcast channel or correlated randomness for this, and quantum cryptography uses a quantum-mechanics-based characteristic that measurement changes the quantum state. How these characteristics can be used has been the subject of extensive discussion. However, long-haul fiber transmissions do not possess such characteristics; another mechanism is needed in this case. We will turn our attention to the fact that the number of the communication parties is limited. For this specific case, we can assume a system wherein Alice and Bob share a common key consisting of random numbers with a uniform distribution (true random numbers) with a fixed length beforehand. The common key could be securely passed by hand, for example. Generally, the common key-sharing method is unprescribed as long as the security is higher than that discussed in this report. The common key gives Bob an advantage over Eve. Here, the common key is not used as a seed key, but is instead used as a tool for transforming the entropy of noise into that of secret keys. For this reason, the information of the common key is not reflected in the information on the transmission channel, and thus, the common key can be repeatedly used. The phase noise of a laser diode's (LD) output, for example, can be used as a noise source (See section 7.4). Phase noise is always present in LD output, and it is sufficiently random [15].

There is a method called the $\alpha\eta$ (Y00) protocol that uses channel noise and a common key [16,17]. This method is basically a stream cipher with quantum fluctuations that uses a common key as a seed key, and it uses multiple bases. However, the method in this report is not a stream cipher, but secret-key generation. Messages are encrypted with the secret keys by one-time pad. A common key is used only inside the transmitter and receiver, and the information on the transmission channel does not reflect the common key-related information. The two methods hence belong to different concepts.

The method in this report is not resistant to brute force attacks because a common key is used. However, the secret keys are continuously generated from noise, and the messages are encrypted by using a one-time pad. Thus, information leaks are restricted in the random-number transmission stage for key agreement. To evaluate the security, Eve is assumed to exactly obtain parity check symbols of an algebraic error-correcting code that is used to transmit the random numbers. Even if the assumption is advantageous to Eve, she must decode the error-correcting code, which is a block code. To do this, she must list the candidates of information symbols by using parity check symbols and list the candidates of the common key. In other words, Eve has no other decrypting method that is more efficient than listing the candidates of information symbols. In addition, the number of candidates increases because of the existence of bit errors, and thus, security is strengthened even more. Computational security is generally achieved by relying on some sort of mathematical difficulty. For example, the security of Diffie-Hellman key agreement is founded on the existence of a difficult computation in number theory [18]. There is no assurance that the difficulty will never be overcome. An efficient algorithm for overcoming that difficulty might be found. However, our method does not assume any mathematical difficulties, and therefore, there is no threat that an efficient decrypting algorithm might be found. The method requires an exhaustive search for the candidates of the common key in decryption. The security of our method does not reach the level of information theoretic security, but it falls into some range of computational security. However, thanks to there being no salient threat, we do not need to be anxious about any unexpected decryptions. Our method will be useful for protecting highly confidential information like government and military secrets.

Cryptography generally has a trade-off between security and convenience. Methods with information-theoretic security have high security but their message transmission rate $R_m$, defined by $R_m = n_m/n_{all}$, where $n_{all}$ is the total number of transmitted bits and $n_m$ is the message part, is low ($R_m \ll 1$), and long-haul transmissions using them are generally difficult. In contrast, methods with computational security achieve $R_m \sim 1$, but generally face the threat that an efficient decrypting method might be found. Supposing we interpret these two kinds of methods as being at opposite ends of a trade-off, our method is located in the middle, because it achieves computational security that removes the threat. However, in so doing, the message transmission rate is reduced to $R_m \ll 1$.

## 2. Framework

This report concerns key agreement consisting of random-number transmission and secret key generation using the transmitted random numbers. Messages are transmitted with a one-time-pad using the generated secret keys. This section describes the framework of the key agreement and defines the security of the method. The notation is such that when a character style like $\mathcal{X}, \mathcal{Y}, \mathcal{Z}$ designates sets, the corresponding random variables are described with capital letters, like $X, Y, Z$, and corresponding elements are described with small letters, like $x, y, z$. Bold letters like $\boldsymbol{x}, \boldsymbol{y}, \boldsymbol{z}$ designate row vectors of $x, y, z$. Letters like $X^n$ designate successive $n$ letters.

Let us assume that there is noise in the transmission channel used for the key agreement.

Therefore, there are generally bit errors in the signals received by Bob and Eve. Let $s_{kA}$ be the secret keys to be shared between Alice and Bob. Generally in a key agreement protocol using a noisy channel, Alice encodes $s_{kA}$ such that $E_g: \{0, 1\}^{n_r} \to \{0, 1\}^n$, $s_{kA} \mapsto x$ and sends them to Bob; he receives and decodes them such that $D_g: \{0, 1\}^n \to \{0, 1\}^{n_r}$, $y \mapsto s_{kB}$ [1-3]. Because of bit errors, generally $x \neq y$. The mapping $E_g$ has two purposes: one is to make the information leaking to Eve meaningless; the other is to achieve accurate communications. The former purpose requires $n_r < n$, and thus $E_g$ is probabilistically performed. However, even if $E_g$ itself is simple, its inverse, i.e., $D_g$, is not easy. Therefore, the probabilistic encoding $E_g$ is not preferable for real systems. Deterministic encoding is better. For this reason, let us invert the process on Alice's side such that $E_g': \{0, 1\}^n \to \{0, 1\}^{n_r}$, $x \mapsto s_{kA}$ [19]. In this case, all processes of the method can be made deterministic and thereby practical. Now, we divide $E_g'$ ($E_g$) into two stages, i.e., the encoding $E$ and the secret key generation $S$, as in Definition 1 below. Figure 1 shows the framework discussed in this report.

As mentioned above, noisy channel models generally have an intrinsic characteristic to make Eve disadvantageous. For example, the broadcast model assumes that Eve suffers from bit errors independent of Bob's ones. However, we do not assume such an intrinsic characteristic. Instead, we assume a common key $k_e$ that Alice and Bob share beforehand to make Eve disadvantageous. The following Definition 1 defines the key agreement protocol discussed in this report.

***Definition 1 [Key agreement protocol]:*** Let us assume that there is noise in the transmission channel. Alice and Bob share beforehand a common key $k_e \in \{0, 1\}^{N_K}$ consisting of a random $N_K$-bit string with uniform distribution over $\{0, 1\}^{N_K}$. Alice encodes a random $n_I$-bit string $x \in \{0, 1\}^{n_I}$ with a uniform distribution over $\{0, 1\}^{n_I}$ by using an $(n, k)$ block code of code length $n$ and information length $k$ ($n > k$), where $x_1 \in \{0, 1\}^k$ are information symbols and $x_2 \in \{0, 1\}^{n-k}$ is redundant information of $x_1$; and she sends $x$ and $x_2$ to Bob. Bob receives $y$ and $y_2$ and obtains $y_1$ that are error-corrected using $k_e$. Alice and Bob respectively generate secret keys $s_{kA}$ and $s_{kB} \in \{0, 1\}^{n_r}$ from $x_1$ and $y_1$.

Encoding $E: \{0, 1\}^{n_I} \times \{k_e\} \to \{0, 1\}^k \times \{0, 1\}^{n-k}$,
$$x \times k_e \mapsto x_1 \times x_2$$
Decoding $D: \{0, 1\}^{n_I} \times \{0, 1\}^{n-k} \times \{k_e\} \to \{0, 1\}^k$,
$$y \times y_2 \times k_e \mapsto y_1$$
Secret key generation $S: \{0, 1\}^k \to \{0, 1\}^{n_r}$,
$$x_1 \mapsto s_{kA} \text{ and } y_1 \mapsto s_{kB}$$

Encoding $E$ uses the common key $k_e$. Decoding $D$ requires $k_e$ and decoding is difficult without $k_e$. A concrete coding method is described in section 5.2. As long as Bob does not fail in decoding, $x_1 = y_1$ and $s_{kA} = s_{kB}$.

In evaluating security, we assume that authentication has been executed and that Eve does not tamper with the channel by inserting or modifying messages. Moreover, we assume that Eve is an outsider.

Let $z$, $z_1$, and $z_2$ be Eve's information corresponding to $x$, $x_1$, and $x_2$ ($y$, $y_1$, and $y_2$) for Alice (Bob). Eve's final aim is to eavesdrop on messages. Because the secret key $s_{kA}$ is used with a one-time pad, Eve needs to derive $s_{kA}$ from $z$ and $z_2$ to achieve her aim. For simplicity, we assume that all $n_r$ bits of $s_{kA}$ is used in message transmissions. Let $z^*$ and $z_2^*$ be another pair of $z$ and $z_2$, and let $s_{kA}^*$ be the secret key generated from $z^*$ and $z_2^*$. If Eve uses a chosen-plaintext attack against the message transmissions, she can obtain any number of $s_{kA}^*$. Here, to simplify the description, all the secret keys that Eve obtains will be represented by $s_{kA}^*$. In accordance with these premises, we assume that Eve's attack and aim are as follows.

***Eve's attack:*** (1) Eve can passively obtain all of the information. (2) Eve can obtain any number of $s_{kA}^*$. (3) Eve cannot control the equipment inside the transmitter and receiver or the environmental noise.

***Attack aim:*** Eve's aim is to guess at least one bit of the secret key $s_{kA}$.

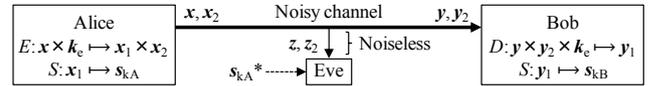

**Fig. 1** Framework discussed in this report.

Because Definition 1 uses a common key, the security of Definition 1 is computational. Thus, we give the following definition.

***Definition 2 [Computational security in key agreement]:*** Let us suppose a game in which Eve runs a probabilistic polynomial algorithm to try to guess one bit at any position of $s_{kA}$ from $z$ ($z^*$), $z_2$ ($z_2^*$), and $s_{kA}^*$ in one arbitrary trial. Let $s_{kE}$ be the guessed one-bit secret key. Let the probability of $s_{kA} = s_{kE}$ be $p_s$, which is called the probability of successfully guessing secret key. A key agreement method is called computationally secure if there exists $k_0 \in \mathbb{N}$ such that $p_s < 1/2 + 1/P(N_K)$ is satisfied at a common key length $N_K \geq k_0$ for every polynomial equation $P(N_K)$.

The security of the framework of Definition 1 is based on the fact that decoding $D$ is difficult without $k_e$. Therefore, let us define the difficulty of decoding along the lines of Definition 2.

***Definition 3 [Computational security in encoding]:*** Let us define an encoding $E_a$ by $x_1 = E_a(x, k_e)$ in accordance with Definition 1. Let us suppose a game in which Eve runs a probabilistic polynomial algorithm to try to guess $x_1$ ($x_1^*$) from $z$ ($z^*$), $z_2$ ($z_2^*$), and $s_{kA}^*$ in one arbitrary trial. Let $z_1$ ($z_1^*$) be the guess of $x_1$ ($x_1^*$). Let the probability of $x_1 = z_1$ ($x_1^* = z_1^*$) be $p_d$, which is called the probability of successfully guessing information symbols. $E_a$ is called computationally secure if there exists $k_0 \in \mathbb{N}$ such that $p_d < 1/2^k + 1/P(N_K)$ is satisfied at a common key length $N_K \geq k_0$ for every polynomial equation $P(N_K)$.

The security of an encryption using a common key is computational; it has no resistance against a brute-force attack on the common key. However, if there is no efficient decrypting method except for a brute-force attack, sufficient security is still obtainable by choosing a sufficiently long key. The issue in computational security is not the lack of resistance against a brute-force attack but the threat that an efficient decrypting method might be found. Therefore, if it is assured that there is no efficient decryption method in an encryption system, the system is sufficiently secure. Definition 2 corresponds to that assurance, but what is assumed in order to achieve the security of Definition 2 is important. If Definition 2 is achieved without assuming any mathematical difficulties such as that in the Diffie-Hellman key agreement, there is no threat that an efficient decryption algorithm might be found. The protocol of Definition 1 ensures the security of Definition 2 by using noise in addition to the common key $k_e$. It does not assume any mathematical difficulties. If the amount of information of secret keys generated in the protocol of Definition 1 is limited to the entropy of noise, the generated secret keys are fresh., Computational security in the sense of Definition 2 can be achieved using this freshness, as will be shown in Theorem 1.

## 3. Concrete method based on Definition 1

The code length $n$ and information symbol length $k$ are generally assumed to be sufficiently long in any key-agreement protocol using noise. However, a practical system might limit the range of $n$ and $k$. Hence, we introduce a parameter $u$ and use $uk$ symbols as a unit of the key generation to overcome the limit. The following Method 1 makes the framework of Definition 1 specific from the viewpoint of an actual system, including the introduction of the parameter $u$. (See Fig. 2).

***Method 1:*** The noise-assisted key-agreement protocol based on Definition 1 consists of algorithms ($R_X$, $E_a$, $E_b$, $S$) in the transmitter and algorithms ($F_t(F_c)$, $D_a$, $D_b$, $S$) in the receiver.

Transmitter:
(1) $x \leftarrow R_X$
(2a) $E_a$: $\{0,1\}^{nl} \times \{k_e\} \to \{0,1\}^k$, $x \times k_e \mapsto x_1$
(2b) $E_b$: $\{0,1\}^k \to \{0,1\}^{n-k}$, $x_1 \mapsto x_2$
(3) $S$: $\{0,1\}^{uk} \to \{0,1\}^{n_r}$, $x_1^u \mapsto s_{kA}$

Receiver:
(1') $y \leftarrow F_t(F_c(x))$
(2a') $D_{1a}$: $\{0,1\}^{nl} \times \{k_e\} \to \{0,1\}^k$, $y \times k_e \mapsto y_1'$
(2b') $D_{1b}$: $\{0,1\}^k \times \{0,1\}^{n-k} \to \{0,1\}^k$, $y_1' \times y_2 \mapsto y_1$
(3') $S$: $\{0,1\}^{uk} \to \{0,1\}^{n_r}$, $y_1^u \mapsto s_{kB}$

Here, $x_1^u$ and $y_1^u$ are respectively $x_1$ and $y_1$ of $u$ blocks. $x_1$ is generated from $x$ with permutations in $E_a$, and $n_l > k$. Let $u \geq 1$, $uk \in \mathbb{N}$, and $uk > n_r$. Here, $n_r \in \mathbb{N}$ is chosen to satisfy $n_r/un \leq C_{s0}$, where $C_{s0}$ is introduced in the next section. This choice makes $E_a$ computationally secure in the sense of Definition 3, as shown in Lemma 11, and makes Method 1 computationally secure in the sense of Definition 2, as shown in Theorem 1. In process (1), $x$ is output from a random-number generator $R_X$ in the transmitter and is transmitted to a receiver. In process (1'), $y$ is received by a receiver, where $y$ includes transmission-carrier noise and environmental noise in $F_c$ and $F_t$, respectively. Here, the symbol "←" is used to show that $x$ is a probabilistic output and $y$ contains a probabilistic bit error. We assume a memory-less binary symmetric channel (BSC) with a BER of $p_E$ as a model of the noise source $F_c$ in the transmitter. An example of system conforming to this model is optical-fiber communications. Light already has fluctuations (noise) that cause bit errors at the moment it is emitted from its source. Another example is noise added on purpose. Processes (2a) and (2b) are the encoding $E$ in Definition 1 and processes (2a') and (2b') are decoding $D$ in Definition 1. $x_1$ is generated from $x$ by using the common key $k_e$ in $E_a$, and $y_1'$ is similarly generated from $y$ by using $k_e$ in $D_a$. A concrete example is described in section 5.2. $E_b$ and $D_b$ are respectively encoding and decoding to achieve errorless communications between Alice and Bob. Thus, $x_1 = y_1$ as long as Bob does not fail in decoding. Redundant information $x_2$ is transmitted through an errorless public channel, and thus, $x_2 = y_2$. Summarizing what has been covered so far, one sees that $x$ and $x_2$ are transmitted from the transmitter and $y$ and $y_2$ are received at the receiver. $x_1$ and $y_1$ are only used inside the transmitter and receiver and they are not transmitted. Processes (3) and (3') describe the secret key generation $S$ in Definition 1, which is achieved through privacy amplification [19,20] using universal hashing. $S$ is performed in units of $u$ blocks. If $x_1^u = y_1^u$, then $s_{kA} = s_{kB}$. Thus, $s_{kA}$ can be shared by Alice and Bob and used in encrypted communications of messages. Figure 2 summarizes the algorithms in the form of a block diagram.

Eve is assumed to be able to receive signals in the best condition; i.e., she receives $z \leftarrow F_c(x)$ for $x$ without environmental noise. Let $p_E$ ($p_E \leq 1/2$) be Eve's bit-error rate, and let $p_B$ ($p_B \leq 1/2$) be Bob's bit-error rate. Because Bob's signals are affected by environmental noise $F_t$, generally $p_B \geq p_E$, where equality corresponds to the case of a noiseless channel.

Redundant information $x_2$ is openly transmitted through an errorless public channel in Method 1. This is because this setup makes the security analysis easy. When $x_2$ is transmitted through a channel with errors, the setup makes Eve disadvantageous. Thus, even if $x_2$ is transmitted through the same channel as $x$ is, the security assured for Method 1 is kept (see section 6.2).

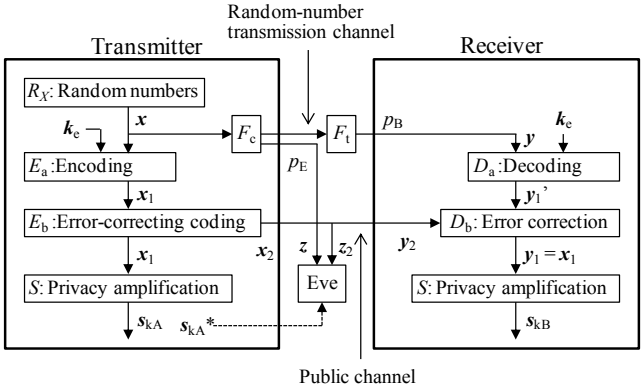

**Fig. 2** Block diagram of Method 1.

## 4. Conditional Secrecy Capacity

The framework of Definition 1 (Method 1) aims to achieve the security of Definition 2 by limiting the entropy of the generated secret keys to that of noise. The idea is the same as that of secrecy capacity [1-4]. However, the secrecy capacity is defined for cases in which secret keys are generated from noise without a common key. It is not defined for the case of using a common key. For this reason, we define the conditional secrecy capacity $C_s$ as a similar quantity. This quantity is defined under the assumption that the common key $k_e$ is secret, and the quantity expresses how many secret keys are generated from noise under the assumption, where "secret" means that Eve has no information about the common key.

The number of bit errors originating from noise statistically fluctuates. $n$ and $k$ should be sufficiently large to reduce the statistical fluctuations in each block. However, their range might be limited in actual systems. Therefore, we define $C_s$ that is applicable even to $n$ and $k$ of limited size by introducing the parameter $u$. The following definition 4 assumes that Alice and Bob share a common key $k_e$ and it is kept secret from Eve. The prerequisite is expressed as " $|k_e \leftarrow K_e$" in the following.

***Definition 4 [Conditional secrecy rate and capacity]:*** In Method 1, the common key $k_e$ is assumed to be kept secret from Eve, and the encoding $E_a$ is assumed to be computationally secure in the sense of Definition 3. Under the assumptions, if the following four conditions are satisfied for a given $\gamma > 0$, $R_s$ is called the conditional secrecy rate for a given $\gamma$. The maximum of $R_s$ is $C_s$ and is called the conditional secrecy capacity for the given $\gamma$.
(1) $\Pr\{S_{kA}{}^{n_r} \neq S_{kB}{}^{n_r} | k_e \leftarrow K_e\} < \gamma$
(2) $\Pr\{I(S_{kA}{}^{n_r} | k_e \leftarrow K_e; S_{kE}{}^{n_r})/n_r < \gamma\} > (1 - \gamma)\{1 - 1/2^k - 1/P(N_K)\}$
(3) $\log_2|S_{kA}{}^{n_r}|/n_r < H(S_{kA}{}^{n_r} | k_e \leftarrow K_e)/n_r + \gamma$
(4) $\Pr\{H(S_{kA}{}^{n_r} | k_e \leftarrow K_e)/un > R_s - \gamma\} > 1 - \gamma$
Here, $S_{kE}{}^{n_r}$ are secret keys of $n_r$ bits generated from $Z^{nl}$ and $Z_2{}^{u(n-k)}$ through Eve's arbitrary guess.

Item (1) assures that Alice and Bob can communicate with each other with a sufficiently small error probability. Item (2) assures that the leaks of secret keys to Eve are sufficiently small. The factor $(1 - \gamma)$ in $\Pr\{\ldots\} > (1 - \gamma)\{1 - 1/2^k - 1/P(N_K)\}$ considers the rare case where the number of bit errors in $un$ symbols is extremely small, owing to statistical fluctuations. The factor $\{1 - 1/2^k - 1/P(N_K)\}$ reflects the assumption that the encoding $E_a$ is computationally secure in the sense of Definition 3. Item (3) assures the uniformity of $S_{kA}{}^{n_r}$. Item (4) indicates the condition that $R_s$ should satisfy in accordance with items (1) – (3). The description $\Pr\{\ldots\} > 1 - \gamma$ considers the rare case where the number of bit errors is extremely small, similar to Item (2).

Reference [19] describes the theory of privacy amplification as methods of generating the secret key $S$. Let Alice's and Bob's information be a random $n_A$-bit string with a uniform distribution over $\{0, 1\}^{n_A}$ and let Eve's corresponding information be $n_E$-bits. Let any $n_s$ of $0 < n_s < n_A - n_E$ be a safety parameter, and let $n_r = n_A - n_E - n_s$. Theorem 3 and corollary 5 of Ref. [19] respectively give $H(S_{kA}{}^{n_r}) = n_r - 2^{n_r - R(X_1{}^{n_A})}/\ln 2$ and $I(S_{kA}{}^{n_r}; S_{kE}{}^{n_r}) \leq 2^{-n_s}/\ln 2$ when Alice and Bob generate an $n_r$-bit string from an $n_A$-bit string by universal hashing [19]. Here, $R(X_1{}^{n_A})$ is the Rényi entropy for the collisions in two independent trials, and it is given by $R(X_1{}^{n_A}) = -\log_2 P_c(X_1{}^{n_A})$ and $P_c(X_1{}^{n_A}) = \sum_{x_{1A} \in \{0,1\}^{n_A}} P_o(x_{1A})^2$ by letting $P_o(x_{1A})$ be the occurrence probability of $x_{1A} \in \{0, 1\}^{n_A}$; $S_{kE}{}^{n_r}$ is the result of Eve's arbitrary guess. Eve's information is $n_E$ bits, but it is not restricted to $n_E$-bit strings.

The claims of Theorem 3 and Corollary 5 in ref. [19] are applicable to Method 1 under the condition that the common key $k_e$ is secret. The parameters $n$ and $k$ in Method 1 are determined such that signals with a bit-error rate of $p_B$ are error-correctable. Let $t_c$ be the number of bit errors definitely correctable per block, which is the lower limit of the maximum number of errors that can be corrected. Let $t_m$ be the upper limit of the maximum number of errors that can be corrected per block, in which bit errors have the possibility of being corrected but the possibility is indefinite. The numbers $t_c$ and $t_m$ are characteristic parameters of the used code.

Definition 4 considers the statistical fluctuations of the bit errors. Now, let us define some quantities as preparation. Let the number of Eve's bit errors per $u$ blocks of information symbols be $n_{ue}$, and let its average be $\bar{n}_{ue} = ukp_E$ and standard deviation be $\sigma_{u2}$. Let $\mathbb{R}_{>0} = \{r \in \mathbb{R} \mid r > 0\}$.

***Lemma 1:*** In Method 1, the common key $k_e$ is assumed to be kept secret from Eve. Let the transmission channel be a memory-less binary symmetric channel (BSC). Let $p_{\sigma E} = p_E \cdot (\bar{n}_{ue} - r\sigma_{u2})/\bar{n}_{ue}$ by using an $r \in \mathbb{R}_{>0}$ that satisfies $\Pr\{n_{ue} < \bar{n}_{ue} - r\sigma_{u2}\} < \gamma$ for a small given $\gamma > 0$. Let $n_s$ be the safe parameter in the secret key generation $S$. If encoding $E_a$ is computationally secure in the sense of Definition 3, the four conditions in Definition 4 can be satisfied by appropriately selecting the parameters $n$, $k$, $u$, and $n_s$ for the small given $\gamma > 0$. The conditional secrecy capacity for the given $\gamma$ is $C_s \geq (k - t_m)/n \cdot h(p_{\sigma E}) - n_s/un$ when using the binary entropy function $h(p) = -p\log_2 p - (1-p)\log_2(1-p)$ (See Fig. 3). The whole secret key $S_{kA}^{n_r}$ is assumed to be used for message transmissions.

***Proof:*** (1) Let $n_{eb}$ be the number of Bob's bit errors per block of code. Let $\varepsilon > 0$ be a parameter that satisfies $1 - (1-\varepsilon)^u \leq \gamma$. The parameters $n$ and $k$ are determined such that $\Pr\{n_{eb} > t_c\} < \varepsilon$ for the small given $\varepsilon > 0$. Bob can generate $Y_1^k$ from $Y^{nl}$ by using the common key $k_e$, and he can correct all the errors except for a small probability $\Pr\{n_{eb} > t_c\} < \varepsilon$. In this case, $\Pr\{S_{kA}^{n_r} \neq S_{kB}^{n_r} \mid k_e \leftarrow K_e\} = 1 - [1 - \Pr\{n_e > t_c\}]^u < 1 - (1-\varepsilon)^u \leq \gamma$. Thus, Definition 4(1) is satisfied.

(2) According to the assumption, the probability with which Eve successfully generates $X_1^k$ without $k_e$ in one arbitrary trial is bounded by $1/2^k + 1/P(N_K)$. Let us suppose that Eve does not succeed in generating $X_1^k$. Even in the case, she obtains $Z^{nl}$ and $Z_2^{n-k}$. Because all of $S_{kA}^{n_r}$ is used in the message transmissions in accordance with the assumption, the $X_1^k$-related information obtainable by Eve is restricted to $Z^{nl}$ and $Z_2^{n-k}$. First, let us consider the information that Eve obtains from only $Z^{nl}$. Because Eve does not have $k_e$, $H(X \mid Z) \leq H((X_1 \mid k_e \leftarrow K_e) \mid Z)$ is satisfied. Here, let "$X_1 \mid k_e \leftarrow K_e$" denote "$X_1$" for simplicity. Then, $H(X \mid Z) \leq H(X_1 \mid Z)$. If Eve's information is only $Z^{nl}$, even though $Z_1^k$ is generated from $Z^{nl}$, the amount of information she gets is unchanged, i.e., $H(X_1 \mid Z) = H(X_1 \mid ZZ_1)$. $H(X_1 \mid ZZ_1) \leq H(X_1 \mid Z_1)$ is generally satisfied. Thus, $H(X \mid Z) \leq H(X_1 \mid Z_1)$. Because $X$ is a binary random number with a uniform distribution, $X_1$ generated from $X^{nl}$ with permutations also has such a property, i.e., $H(X) = H(X_1) = 1$. Thus, $I(X;Z) = H(X) - H(X \mid Z) \geq H(X_1) - H(X_1 \mid Z_1) = I(X_1;Z_1)$. Next, let us consider the information that Eve obtains from $Z_2^{n-k}$ as well as $Z^{nl}$. When $Z_1^k$, a permutation of $Z^{nl}$, and the $Z_2^{n-k}$ function as a code, $Z_1^k$ is error-corrected and Eve obtains $X_1^k$. This case is included in the case in which Eve succeeded in generating $X_1^k$. Because we are discussing the case in which Eve does not succeed in generating $X_1^k$, $Z_1^k$ and $Z_2^{n-k}$ do not function as a code. In this case, Eve cannot correct errors, but $Z_2^{n-k}$ involves redundant information for correcting $t_m$ bits of the errors of $Z_1^k$ at maximum, where $t_m$ is a characteristic parameter of the used code. If the function of $Z_2^{n-k}$ is evaluated most advantageously from the Eve's standpoint, the effect of $Z_2^{n-k}$ is to repair $H(X_1) - H(X_1 \mid Z_1)$ back to $H(X_1)$ for $t_m$ symbols of $Z_1^k$ at maximum (see Fig. 3(c)). For the remaining $(k - t_m)$ symbols, the mutual information $H(X_1) - H(X_1 \mid Z_1)$ is unchanged because of the correction limit of the used code. Hence, when Eve does not succeed in generating $X_1^k$, the amount of information per $u$ blocks is

$n_E \leq u t_m H(X_1) + u(k - t_m)[H(X_1) - H(X_1 \mid Z_1)]$
$\leq u t_m H(X_1) + u(k - t_m)[H(X) - H(X \mid Z)]$.

Because $H(X) = H(X_1) = 1$,

$n_E \leq u t_m + u(k - t_m)[1 - H(X \mid Z)]$.

$H(X \mid Z)$ is $H(X \mid Z) = h(p_E)$ using the average bit-error rate. However, if the actual number of bit errors in one block is less than the average number of bit errors determined by $p_E$, Eve actually obtains more information than the average amount of information. Therefore, we must take Eve's situation into account by considering the statistical fluctuations of the bit errors. In particular, we will consider the statistical fluctuations for $u$ blocks in the bit sequence because the unit of the secret key generation is $u$ blocks. Because $H(X \mid Z)$ is described using a bit-error rate, we describe the statistical fluctuations by using those of the bit-error rate that is evaluated for every $u$ blocks of the bit sequence. Because $p_{\sigma E} = p_E \cdot (\bar{n}_{ue} - r\sigma_{u2})/\bar{n}_{uE}$ is defined using $r$ that satisfies $\Pr\{n_{ue} < \bar{n}_{ue} - r\sigma_{u2}\} < \gamma$, $H(X \mid Z)$ in each sequence of $u$ blocks satisfies $H(X \mid Z) \geq h(p_{\sigma E})$ except for a small probability $\Pr\{n_{ue} < \bar{n}_{ue} - r\sigma_{u2}\} < \gamma$. In this case,

$n_E \leq u t_m + u(k - t_m)[1 - h(p_{\sigma E})]$.

Because Alice's information per $u$ blocks is $n_A = uk$, we have

$n_A - n_E \geq u(k - t_m)h(p_{\sigma E})$.

Let $n_r = n_A - n_E - n_s$ for any positive safe parameter $n_s < n_A - n_E$. According to Corollary 5 in ref. [19], $I(S_{kA}^{n_r} \mid k_e \leftarrow K_e; S_{kE}^{n_r}) \leq 2^{-n_s}/\ln 2$ can be achieved by universal hashing. Because of $n_A - n_E = O(u)$,[2] $n_s$ and $n_r$ can be also chosen to be $O(u)$. Thus, $I(S_{kA}^{n_r} \mid k_e \leftarrow K_e; S_{kE}^{n_r})/n_r \leq 2^{-n_s}/n_r \ln 2 < \gamma$ can be satisfied for the given $\gamma$ by appropriately choosing $u$. This relation is satisfied except for the small probability $\Pr\{n_{ue} < \bar{n}_{ue} - r\sigma_{u2}\} < \gamma$ and for the case that Eve does not succeed in generating $X_1^k$. $p_d$ in Definition 3 is the probability of successfully guessing information symbols for one block. Let $p_{du}$ be this probability for $u$ blocks. Generally, $1 - p_{du} \geq 1 - p_d$, and from the assumption, $1 - p_d > 1 - 1/2^k - 1/P(N_K)$. Thus, $1 - p_{du} > 1 - 1/2^k - 1/P(N_K)$. According to the above-mentioned two conditions, $\Pr\{I(S_{kA}{}^{n_r}|k_e \leftarrow K_e; S_{kE}{}^{n_r})/n_r < \gamma\} > (1-\gamma)(1-p_{du})$. Hence, $\Pr\{I(S_{kA}{}^{n_r}|k_e \leftarrow K_e; S_{kE}{}^{n_r})/n_r < \gamma\} > (1-\gamma)\{1 - 1/2^k - 1/P(N_K)\}$, and Definition 4(2) is satisfied.

---

[2] In this report, notations $O(u)$ and $O(1/u)$ are used for $u \to \infty$.

(3) According to Theorem 3 in Ref. [19], $H(S_{kA}{}^{n_r}|k_e \leftarrow K_e) \geq n_r - 2^{n_r - R(X_1^{n_A})}/\ln 2$ is obtained. Definition 1 assumes that $P_o(x_{1A})$ has a uniform probability, and thus $P_c(X_1^{n_A}) = \sum_{x_{1A} \in \{0,1\}^{n_A}} P_o(x_{1A})^2 = 2^{-n_A}$ and $R(X_1^{n_A}) = -\log_2 P_c(X_1^{n_A}) = n_A$. Thus, $H(S_{kA}{}^{n_r}|k_e \leftarrow K_e) \geq n_r - 2^{n_r - n_A}/\ln 2$. Because $n_r - n_A = -n_E - n_s$, and $n_s$ and $n_r$ are chosen to satisfy $2^{-n_s}/n_r \ln 2 < \gamma$, the relation $H(S_{kA}{}^{n_r}|k_e \leftarrow K_e)/n_r \geq 1 - 2^{-n_E - n_s}/n_r \ln 2 \geq 1 - 2^{-n_s}/n_r \ln 2 > 1 - \gamma$ is obtained. Because of $|S_{kA}{}^{n_r}| = n_r$, $\log_2|S_{kA}{}^{n_r}|/n_r = 1$ is obtained. Thus, $H(S_{kA}{}^{n_r}|k_e \leftarrow K_e)/n_r > \log_2|S_{kA}{}^{n_r}|/n_r - \gamma$ is satisfied. Hence, Definition 4(3) is satisfied.

(4) Using $H(S_{kA}{}^{n_r}|k_e \leftarrow K_e) \geq n_r - 2^{n_r - n_A}/\ln 2$, $n_r = n_A - n_E - n_s$, and $n_A - n_E \geq u(k - t_m)h(p_{\sigma E})$, which is satisfied except for the small probability $\Pr\{n_{ue} < \bar{n}_{ue} - r\sigma_{u2}\} < \gamma$, we obtain

$H(S_{kA}{}^{n_r}|k_e \leftarrow K_e)/un + \gamma$
$\geq n_r/un - 2^{n_r - n_A}/un \ln 2 + \gamma$
$\geq (k - t_m)/n \cdot h(p_{\sigma E}) - n_s/un + (\gamma - 2^{-n_s - n_E}/un \ln 2)$.

Using $\gamma > 2^{-n_s}/n_r \ln 2 > 2^{-n_s}/un \ln 2 > 2^{-n_s - n_E}/un \ln 2$, we obtain $(\gamma - 2^{-n_s - n_E}/un \ln 2) > 0$. Therefore, $H(S_{kA}{}^{n_r}|k_e \leftarrow K_e)/un + \gamma > (k - t_m)/n \cdot h(p_{\sigma E}) - n_s/un$. Definition 4 (4) requires $H(S_{kA}{}^{n_r}|k_e \leftarrow K_e)/un + \gamma > R_s$ except for the small probability $\Pr\{n_{ue} < \bar{n}_{ue} - r\sigma_{u2}\} < \gamma$. If $R_s = (k - t_m)/n \cdot h(p_{\sigma E}) - n_s/un$ is selected, it satisfies Definition 4 (4). As long as $R_s$ is less than that, Definition 4 (4) is satisfied. Therefore, the selected value is the lower bound of $C_s$, where $C_s$ is the maximum of $R_s$. Thus, if $n$, $k$, $u$, and $n_s$ are appropriately selected in accordance with the above discussion, the conditional secrecy capacity for the given $\gamma$ is $C_s \geq (k - t_m)/n \cdot h(p_{\sigma E}) - n_s/un$. □

Let $C_{s0}$ be the lower bound of $C_s$ in Lemma 1, i.e., $C_{s0} = (k - t_m)/n \cdot h(p_{\sigma E}) - n_s/un$. The conditional secrecy capacity originates from the entropy $h(p_{\sigma E})$ of bit errors, as shown in Fig. 3. The common key is used only for transforming the entropy $h(p_{uE})$ of bit errors into that of secrecy keys. Therefore, the conditional secrecy capacity maintains $C_s > 0$ for repeated use of $k_e$. Method 1 restricts the secret key-generation rate to $C_{s0}$ in order to repeatedly use the common key $k_e$. Lemma 1 assumes that Method 1 is computationally secure in the sense of Definition 3, and it is in Lemma 11 that security is proved. The reason why Lemma 1 is shown here prior to Lemma 11 is to determine the amount of $S_{kA}{}^{n_r}$, i.e., $n_r \leq unC_{s0}$.

For simplicity, Lemma 1 assumes that all of $S_{kA}{}^{n_r}$ is used in message transmissions. On the other hand, the case in which only part of $S_{kA}{}^{n_r}$ is used in message transmissions is as follows. For example, when $n_r'$ bits are used in message transmissions and $(n_r - n_r')$ bits are leaked to Eve, $C_{s0}$ is transformed into $C_{s0}' = (unC_{s0} - n_r + n_r')/un$. Although the conditional secrecy capacity varies depending on the amount of leaked information, the fact that $unC_{s0}'$ indicates the capacity actually needed in message transmissions does not change. For this reason, Lemma 1 assumed that all of $S_{kA}{}^{n_r}$ is used in message transmissions.

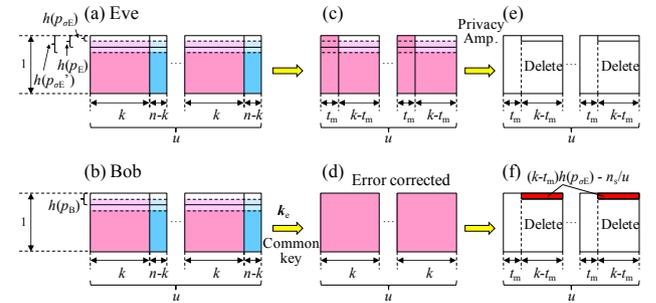

**Fig. 3** Amount of information in the key-agreement protocol. Colored areas indicate information that Eve and Bob have. Faded colors indicate the range of statistical fluctuations. Blank areas indicate no information. $u$ is the unit of secret key generation. (a) Eve's received amount of information, evaluated in terms of the BER of $p_E$ information theoretically. Here, $p_{\sigma E} = p_E \cdot (\bar{n}_{uE} - r\sigma_{u2})/\bar{n}_{uE}$ and $p_{\sigma E}' = p_E \cdot (\bar{n}_{uE} + r\sigma_{u2})/\bar{n}_{uE}$. (b) Bob's received amount of information. (c) Eve's amount of information after considering the redundant information. (d) Bob can correct errors using the common key $k_e$. (e), (f) Eve's amount of information is deleted in the privacy amplification, where the statistical fluctuations of bit errors are considered. The remaining amount of information is secret. The term related to the safe parameter $n_s$ is not drawn because it is so small.

## 5. Coding

The conditional secrecy capacity in Lemma 1 includes a parameter $t_m$ that is the upper limit of the maximum number of errors that can be corrected per block. Therefore, we need to clarify $t_m$ in Method 1. For this purpose, an $(n, k)$ linear code like Reed-Solomon (RS) code can be used [22,23]. We cannot use recent high-performance codes that use "probabilistic" characteristics like Low Density Parity Check (LDPC) code instead of algebraic codes [22,23], because their performance is near the Shannon limit, and the upper limit of their maximum number of errors that they can correct is not definite.

## 5.1 Maximum Number of Errors Correctable by an Algebraic Code

The upper limit of the maximum number of code word errors $t_{mc}$ that an algebraic code can correct is determined by the Hamming distance between code word vectors. Here, the algebraic code is not limited to a binary code, and $t_{mc}$ is defined for a general code. When the coding is binary, $t_{mc} = t_m$, and when it is over $GF(2^m)$, $t_m = mt_{mc}$. There is a theorem called the Singleton bound for an $(n, k)$ linear code, i.e., $d \leq n - k + 1$, where $n$ is the code word length, $k$ is information symbol length, and $d$ is minimum distance [22,23]. When equality is satisfied in this theorem, the corresponding code is called a maximum distance separable (MDS) code. Reed-Solomon is the most practical such code. When the Hamming weight $w_H(e)$ of an error $e$ is $w_H(e) \leq (d-1)/2$, the error can be exactly corrected. This is a classical bound in error correction. When $(d-1)/2 < w_H(e) \leq d-1$, the candidates for the code word vectors can be listed, and the error has the possibility of being corrected (list decoding). However, when $w_H(e)$ is beyond $d-1$, the code word vector with the error usually enters the region of another code word vector and the error is not correctly detected. MDS codes have this characteristic for almost all errors, and the upper limit of the maximum number of errors that the codes can correct is given by the distance $t_{mc} = d-1$. This distance is equal to $n-k$ in MDS codes, i.e., $t_{mc} = n-k$, and this is intuitively understandable because $n-k$ is the number of redundant code words. The estimate of $t_{mc} = n-k$ for the upper limit of the maximum number of errors that the codes can correct has a sufficient margin, because although recent studies have shown the possibility of list decoding [24,25], correctability is restricted to the relatively nearby region of $(d-1)/2$ for practical choices of $n$ and $k$.

The above paragraph describes the case of hard-decision decoding. There is also soft-decision decoding. However, soft-decision decoding extends the classical bound only by one or a few code words depending on the code employed [26,27]. This quantity is sufficiently small compared with the $t_{mc} = d-1$ bound described above for list decoding for sufficiently large $d$.

As described above, MDS codes are excellent from the viewpoint of clarifying the upper limit of the maximum number of errors $t_{mc}$ that the codes can correct. For that reason, any practical system would use MDS codes. The example shown in sections 6.1.2 and 7.2 is a case of using MDS codes.

## 5.2 Concrete coding method

This section describes a concrete example of the encoding $E_a$ and $E_b$. Encoding $E_a$ divides $x^{(N_K)} \in \{0, 1\}^{N_K}$ into $b_I^{(N_1)} \in \{0, 1\}^{N_1}$ and $b_{II}^{(N_2)} \in \{0, 1\}^{N_2}$ by using $k_e \in \{0, 1\}^{N_K}$.[3] $E_b$ encodes $b_I$ and $b_{II}$ independently by using an $(n, k)$ linear code over $GF(2^m)$. Here, $N_1$ and $N_2 \in \mathbb{N}$ satisfy $N_1 + N_2 = N_K$. The following is a concrete example of $E_a$ and $E_b$.

***Coding 1*** *[with common key]:*

$E_a$: $\{0, 1\}^{N_K} \times \{k_e\} \rightarrow \{0, 1\}^{N_1} \times \{0, 1\}^{N_2}$,

   $x^{(N_K)} \times k_e \mapsto b_I^{(N_1)} \times b_{II}^{(N_2)}$,

   where $x \rightarrow b_I$ for $k_e = 1$ and $x \rightarrow b_{II}$ for $k_e = 0$.

$E_b$ [Systematic $(n, k)$ coding over $GF(2^m)$]:

   $\{0, 1\}^{mk} \rightarrow \{0, 1\}^{m(n-k)}$, $b_I \mapsto c_I$ and $b_{II} \mapsto c_{II}$

Here, $c_I$ and $c_{II}$ are respectively parity check symbol vectors of $b_I$ and $b_{II}$. $N_1$ and $k$ satisfy $-r_1\sigma_1 \leq N_1 - \overline{N_0} \leq r_1\sigma_1$ and $\overline{N_0} + \lfloor r_1\sigma_1 \rfloor \leq mk$ for $r_1 \in \mathbb{R}_{>0}$, typically $r_1 = 3$, where $\overline{N_0} = N_K/2$ and $\sigma_1^2 = N_K/4$. $k_e$ is repeatedly used.

The above restrictions on $N_1$ and $k$ are to prevent Eve from deriving the common key $k_e$ part-by-part, as will be described in section 5.3. $N_1$ and $N_2$ respectively denote the numbers of "1"s and "0"s in $k_e \in \{0, 1\}^{N_K}$. Figure 4 schematically shows Coding 1. The random number sequence $x \in \{0, 1\}^{nl}$ is divided into two groups depending on "0" and "1" in $k_e$. The first bit of $k_e$ is "1" in Fig. 4, and the first bit of $x$ is allocated to group I. The second bit of $k_e$ is "0," and the second bit of $x$ is allocated to group II. Subsequent bits are similarly allocated. Random numbers in each group are error-correcting coded independently group-by-group. Because the coding is group-by-group, if the grouping is not correctly done in the receiver, parity check symbols cannot be used. Because Eve does not know the common key, she cannot divide the random number sequence into groups or correct the bit errors. This impossibility makes secret communications possible.

The common key needs to be extended to handle a long random number sequence $x$. However, we will simply use $k_e$ repeatedly to evaluate the basic performance of this method. Of course, were there an extension that used $k_e$ as a seed key of pseudo-random numbers, its cryptographic power would be computationally strengthened.

---

[3] The reason why the notations $x^{(N_K)}$, $b_I^{(N_1)}$, and $b_{II}^{(N_2)}$ are introduced is to differentiate them from $x \in \{0, 1\}^{nl}$ and $b_I$ & $b_{II} \in \{0, 1\}^{mk}$, respectively.

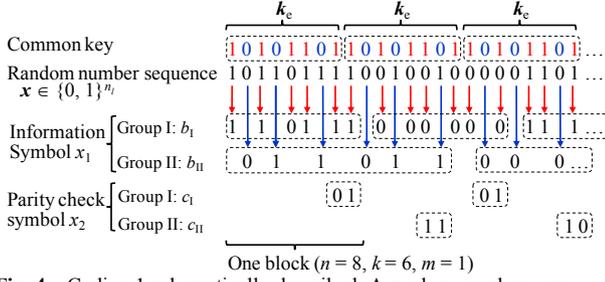

**Fig. 4** Coding 1 schematically described. A random number sequence is divided into two groups in accordance with the "0s" and "1s" in the common key. Each group is independently error-correcting coded. In this figure, the key length is $N_K = 8$, the information symbol length is $k = 6$, the parity check symbol length is $n - k = 2$, and $m = 1$.

### 5.3 Common Key Length and Code Length

Although Definition 1 (Method 1) uses a common key $k_e$, the random number sequence $x$ itself consists of true random numbers, and it never reflects $k_e$. However, because Eve can get redundant information $z_2 = x_2$ (the parity check symbols $c_I$ and $c_{II}$ in Coding 1), she can possibly derive $k_e$ from it. The restrictions on $N_1$ and $k$ imposed in Coding 1 are to minimize this possibility. In this section, we discuss these restrictions.

***Lemma 2:*** In Coding 1 using a common key $k_e$, if $\max(N_1, N_2) \leq mk$, then all information of the common key $k_e$ is needed even when coding one block.

***Proof:*** When each symbol of $x$ of $N_K$ bits is allocated to $b_I$ or $b_{II}$ using $k_e$ in accordance with Coding 1, if $\max(N_1, N_2) \leq mk$, the numbers of $b_I$ and $b_{II}$ are less than or equal to $mk$. Therefore, all information of the common key is needed even when coding one block. □

When a block code is used, the $k_e$-deriving process using parity check symbols must be performed in units of one block. If $\max(N_1, N_2) > mk$, there is part of $k_e$ that is not used for forming one block, and deriving part of $k_e$ becomes possible. Therefore, $N_K$ and $k$ should be determined under the condition of $\max(N_1, N_2) \leq mk$. Let us describe $N_1$ and $N_2$ as functions of $k_e$, i.e., $N_1(k_e)$ and $N_2(k_e)$. If we wholly consider $\{0, 1\}^{N_K}$ as $k_e$, $\max_{k_e}(N_1(k_e), N_2(k_e)) = N_K$. However, many cases satisfy $N_1(k_e) \sim N_2(k_e) \sim N_K/2$; therefore, we will restrict the set of common keys to the case satisfying $N_1(k_e) \sim N_2(k_e) \sim N_K/2$ as follows.

*[Set of common keys $k_e$]:* The set $\mathcal{K}_e$ of common keys $k_e$ of length $N_K$ is restricted to

$\mathcal{K}_e = \{ k_e \in \{0, 1\}^{N_K} \mid -r_1\sigma_1 \leq N_1(k_e) - \overline{N_0} \leq r_1\sigma_1 \}$.

Here, $r_1$ is a design parameter that is typically chosen to be 3.

Because $N_1(k_e) + N_2(k_e) = N_K \in \mathbb{N}$, if $-r_1\sigma_1 \leq N_1(k_e) - \overline{N_0} \leq r_1\sigma_1$, then automatically $-r_1\sigma_1 \leq N_2(k_e) - \overline{N_0} \leq r_1\sigma_1$. Therefore, if $k_e \in \mathcal{K}_e$, then $\max_{k_e}(N_1(k_e), N_2(k_e)) = \overline{N_0} + \lfloor r_1\sigma_1 \rfloor$, where $\lfloor x \rfloor$ denotes the maximum integer $\leq x$. Thus, if $k$ is determined according to $\overline{N_0} + \lfloor r_1\sigma_1 \rfloor \leq mk$, all of $k_e$ is used to form one block in accordance with Lemma 2. The restrictions imposed on $N_1$ and $k$ in Coding 1 are for the above reasons.

$\Pr\{k_e' \in \mathcal{K}_e\}$ for $k_e' \in \{0, 1\}^{N_K}$ is estimated as follows. The probability that each bit of a randomly chosen $k_e'$ is 0 or 1 is $p = 1/2$. Thus, $N_j(k_e')$ ($j = 1$ and $2$) obeys a binomial distribution $P(N_j(k_e')) = \binom{N_K}{N_j(k_e')} p^{N_j(k_e')}(1-p)^{N_K - N_j(k_e')}$. The average is $\overline{N_0} = N_K/2$, and the variance is $\sigma_1^2 = N_K p(1-p)$. Thus, $\Pr\{k_e' \in \mathcal{K}_e\} = \sum_{N_j = \overline{N_0} - \lceil r_1\sigma_1 \rceil}^{\overline{N_0} + \lfloor r_1\sigma_1 \rfloor} P(N_j)$, where $\lceil x \rceil$ denotes the minimum integer $\geq x$. Let $\delta = \Pr\{k_e' \in \{0, 1\}^{N_K} \setminus \mathcal{K}_e\}$. $\delta$ is given by $\delta = 1 - \Pr\{k_e' \in \mathcal{K}_e\}$. For example, when $r_1 = 3$ and the binomial distribution is approximated with a normal distribution, $\Pr\{k_e' \in \mathcal{K}_e\} = 0.9973$, and $\delta = 0.0027$.

### 6. Deriving the common key

The process by which Eve tries to derive the common key $k_e$ is equivalent to her trying to derive $x_1$ ($x_1^*$) of one block as described in this section. To derive secret key $s_{kA}$ in Method 1, $u$ blocks of $x_1$ are needed. Therefore, the computational complexity of deriving $s_{kA}$ is at least that of deriving $k_e$ as shown in Lemma 10. Thus, we first evaluate the computational complexity of deriving $k_e$. The information obtainable by Eve is $z$ ($z^*$), $c_I$ ($c_I^*$), $c_{II}$ ($c_{II}^*$), and $s_{kA}^*$ that is not used in message transmissions, where $c_I^*$ and $c_{II}^*$ are the parity check symbol vectors corresponding to $z_2^*$. First, we will consider that only $z$ ($z^*$), $c_I$ ($c_I^*$), and $c_{II}$ ($c_{II}^*$) are leaked and estimate the computational complexity of deriving $k_e$. Section 6.1.1 considers the case without bit errors, and section 6.1.2 considers the case with bit errors. Next, section 6.2 takes $s_{kA}^*$ into consideration, and it is shown that the computational complexity of deriving $k_e$ does not decrease even if $s_{kA}^*$ is taken into consideration (Lemma 8). Using these results, the computational complexity of deriving $k_e$ is quantified (Lemma 9), and Method 1 using Coding 1 is proved to be computationally secure in the sense of Definition 2 (Theorem 1).

## 6.1 Deriving the common key using parity check symbols

### 6.1.1 Case without bit errors

Because Eve can obtain the parity check symbols $c_I$ and $c_{II}$ in Method 1, if $x$ ($y$) is transmitted without bit errors ($z = x$), she can derive $k_e$. Let us estimate the computational complexity of deriving $k_e$. The routine of Coding 1 is "$x^{(N_K)} \times k_e \mapsto b_I^{(N_1)} \times b_{II}^{(N_2)}, b_I \times b_{II} \mapsto c_I \times c_{II}$." The information that Eve can obtain is $z$ and $c_I \times c_{II}$. Because $z$ itself has no information, the derivation of $k_e$ is based on $c_I \times c_{II}$. Here, $b_I \times b_{II}$ are derived from $c_I \times c_{II}$, and then $k_e$ is derived by comparing $b_I \times b_{II}$ and $z$.

Figure 5 shows the relation between the random number sequence and the first block of group I. Let $\mathcal{G}$ be the set of all elements over $GF(2^m)$. Let $b_0 \in \mathcal{G}^k$ be an information symbol vector in the first block of group I that is obtained from a random number sequence $x$ using $k_e$. We will describe $x \mapsto b_0$ as $b_0 = f(x|k_e)$, where $b_0$ is a row vector with $k$ components over $GF(2^m)$. Let $c^{(p0)} \in \mathcal{G}^{n-k}$ be the parity check symbol vector corresponding to $b_0$. $c^{(p0)}$ is given by $c^{(p0)} = b_0 G_p$, where $G_p$ is the parity check symbol generating part of the generator matrix $G$. The following lemma states a quantitative property about $c^{(p0)}$.

***Lemma 3:*** When only a parity check symbol vector is given in an $(n, k)$ linear code over $GF(2^m)$, $2^{mk}/2^{m(n-k)}$ kinds of information symbol vectors exist for each parity check symbol vector.

***Proof:*** An $(n, k)$ linear code over $GF(2^m)$ consists of $mk$ bits of information symbols and $m(n - k)$ bits of parity check symbols. When the information symbols are derived from only parity check symbols, $mk - m(n - k)$ bits cannot be determined. Therefore, $2^{mk-m(n-k)}$ kinds of information symbol vectors exist for each parity check symbol vector. □

Let us define the set $\mathcal{B}_0$ for the $2^{mk}/2^{m(n-k)}$ kinds of information symbol vectors that are associated with $c^{(p0)}$:

$\mathcal{B}_0 = \{ b_0' \mid c^{(p0)} = b_0' G_p \}$.

Of course, $b_0 \in \mathcal{B}_0$.

Next, we define the set $\mathcal{K}_{e0}$ by using $\mathcal{B}_0$ and $\mathcal{K}_e$:

$\mathcal{K}_{e0} = \{ k_e' \in \mathcal{K}_e \mid b_0' = f(x|k_e') \in \mathcal{B}_0 \}$.

The elements of $\mathcal{K}_{e0}$ are the candidates of the common key. The number of candidates can be determined as follows:

***Lemma 4:*** Suppose a common key $k_e \in \mathcal{K}_e$ is used according to Coding 1. A random number sequence and a parity check symbol vector for the first block of group I or II are exactly given, and one of the positions of the random number sequence corresponding to the first bit of the common key is given to form the first block. The number of the candidates for the common key in this case is $N_{cand} = 2^{N_K - m(n-k)}(1-\delta)$ on average. Here, $(1-\delta)$ is a factor due to $k_e \notin \{0, 1\}^{N_K} \backslash \mathcal{K}_e$.

***Proof:*** Let the random number sequence be $x$, and let the parity check symbol vector be $c^{(p0)}$, where the parity check symbol vector is represented by that of the first block of group I. The candidates of the common key are obtained by listing the elements of $\mathcal{B}_0$, comparing the elements with $x$, and listing the elements of $\mathcal{K}_{e0}$. The parity check symbol vectors are of $2^{m(n-k)}$ kinds, and the number of elements of $\mathcal{K}_e$ is $2^{N_K}(1-\delta)$. In this case, when a parity check symbol vector $c^{(p0)}$ is given, the number of candidates of the common key is $N_{cand} = 2^{N_K - m(n-k)}(1-\delta)$ on average. □

The information obtainable by Eve about the first block of group I is $z = x$ and $c^{(p0)}$. According to Lemma 4, Eve can narrow down the candidates of $k_e$ to $N_{cand}$ on average. This number can be made tremendously large if we appropriately choose $N_K$, $m$, $n$, $k$, and $\delta$. However, a listing is possible in principle even though no memory with a high enough capacity exists. Eve can check each of the listed elements by decoding the blocks of group II and other blocks of group I, and she can continue this process until the candidates of $k_e$ have been narrowed down to one.

***Corollary 1:*** Let us assume that only a random number sequence and parity check symbols are given in Method 1 using Coding 1. It is impossible to derive only part of the common key.

***Proof:*** This claim is apparently true from the fact that deriving the common key is processed in units of one block and that one block is constructed using all the information about the common key, due to the condition $\overline{N_0} + \lfloor r_1 \sigma_1 \rfloor \leq mk$. □

***Corollary 2:*** Let us assume that only a random number sequence and parity check symbols are given in Method 1 using Coding 1. The computational complexity of deriving at least one bit of the common key is $O(N_{cand})$ under the condition that no bit errors exist. In other words, an exhaustive search of $N_{cand}$ is needed.

***Proof:*** Because deriving only part of the common key is impossible according to Corollary 1, the whole common key needs to be derived even for only one bit. In this case, the computational complexity is $O(N_{cand})$ because the process in narrowing down the candidates of the common key based on Lemma 4 involves the complexity of $O(N_{cand})$. □

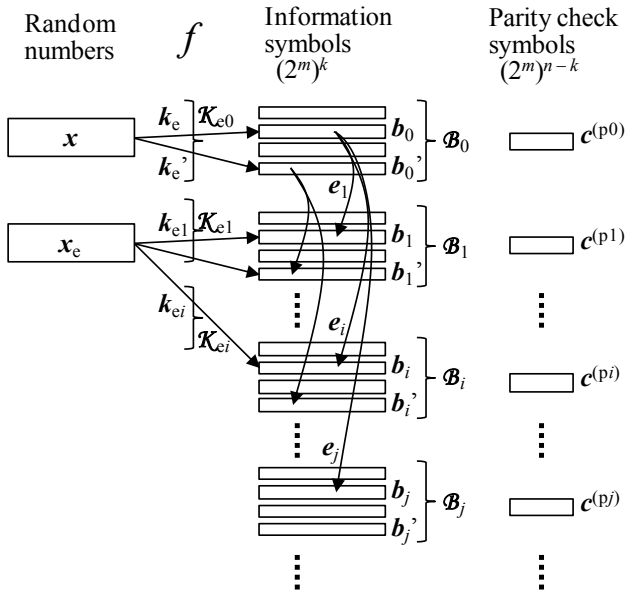

**Fig. 5** Relation between random number sequence and information symbols + parity check symbols in the first block of group I. Here, $z = x_e$.

6.1.2 Case with bit errors

When bit errors exist, Eve needs to consider all possible error patterns to derive the common key. She modifies information symbols according to each bit-error pattern and uses the strategy described in the above subsection.

　　　The number of errors obeys a binominal distribution, and the occurrence probability is highest for the average number of errors. Therefore, let us first concretely estimate the number of error patterns for the average number of errors.

　　　There is no bit error in parity check symbols in Method 1. The number of information symbols in an $(n, k)$ code over $GF(2^m)$ is $mk$ bits, and the average number of bit errors is $\bar{n}_e = p_E mk$ in one block. The number of bit-error patterns is $\binom{mk}{\bar{n}_e}$, and it can be approximated using Stirling's formula, $n! \simeq \sqrt{2\pi n}\, n^n e^{-n}$, as

$$\binom{mk}{\bar{n}_e} \simeq \sqrt{\frac{mk}{2\pi(mk-\bar{n}_e)\bar{n}_e}} \left(\frac{mk}{mk-\bar{n}_e}\right)^{mk-\bar{n}_e}\left(\frac{mk}{\bar{n}_e}\right)^{\bar{n}_e}.$$

　　　The parameters $n$ and $k$ are determined so as to correct errors with a sufficient margin; e.g., the $3\sigma_2$ region is included inside the error-correctable region with a sufficient margin, where $\sigma_2$ is the standard deviation of the bit-error distribution. An MDS code is error-correctable for code word errors satisfying $w_H(e) \leq (n-k)/2$, as described in section 5.1; therefore, the bit errors up to $(n-k)/2$ are exactly correctable, and $\bar{n}_e + 3\sigma_2 < (n-k)/2$ is the condition for determining $k$. For example, Let us consider the case of $m = 8$, $n = 2^m - 1 = 255$, $k = 167$, and $p_{eff} = 0.1$, where $p_{eff}$ is the code-error rate given by $p_{eff} = 1 - (1 - p_E)^m$. In this case, $\bar{n}_e \simeq 17.5$, $\sigma_2 \simeq 4.15$, $\bar{n}_e + 3\sigma_2 = 29.95 < (n-k)/2 = 44$, and $\binom{mk}{\bar{n}_e} \simeq 2.8 \times 10^{39}\ (1.0 \times 2^{131})$.

***Lemma 5:*** Let us assume that only a random number sequence and parity check symbols are given in Method 1 using Coding 1. Let $N_{ep}$ be the number of error patterns in one block. When bit errors exist, the computational complexity of deriving at least one bit of the common key is $O(N_{ep} \cdot N_{cand})$.

***Proof:*** Let us choose the first block of group I as a representative (see Fig. 5). Let $e_i$ be an error vector in the first block of group I, where the Hamming weights satisfy $w_H(e_i) \leq (d-1)/2$, $i = 1, 2, 3, \ldots$ We define $b_i' = b_0' + e_i$ for $b_0' \in \mathcal{B}_0$. Given $b_i = b_0 + e_i$ and $c^{(pi)} = b_i G_p$, then $b_i' G_p = (b_0' + e_i) G_p = (b_0' + b_i - b_0) G_p = b_i G_p = c^{(pi)}$, and $\mathcal{B}_0 \cap \mathcal{B}_i = \phi$ for $\mathcal{B}_i = \{b_i' \mid c^{(pi)} = b_i' G_p\}$. Let $x_e$ be a random number sequence with bit errors. Moreover, if we define $\mathcal{K}_{ei} = \{k_{ei} \in \mathcal{K}_e \mid b_i' = f(x_e|k_{ei}) \in \mathcal{B}_i\}$, the elements of $\mathcal{K}_{ei}$ are the candidates of the common key in error pattern $e_i$. Similarly, if we define $b_j = b_0 + e_j$, $c^{(pj)} = b_j G_p$, $\mathcal{B}_j = \{b_j' \mid c^{(pj)} = b_j' G_p\}$, and $\mathcal{K}_{ej} = \{k_{ej} \in \mathcal{K}_e \mid b_j' = f(x_e|k_{ej}) \in \mathcal{B}_j\}$ for another error vector $e_j$ ($w_H(e_j) \leq (d-1)/2$, $j \neq i$), then $\mathcal{B}_0 \cap \mathcal{B}_j = \phi$ and $\mathcal{B}_i \cap \mathcal{B}_j = \phi$. Because $b_i' = f(x_e|k_{ei})$ is a single-valued function, $\mathcal{K}_{ei} \cap \mathcal{K}_{ej} = \phi$. Thus, there is no overlap between the candidates of the common key for different error vectors. Because the occurrence of each error pattern is probabilistic, Eve must consider all such patterns when deriving the common key and there are candidates of the common key described in Lemma 4 for every pattern. Thus, the computational complexity of deriving at least one bit of the common key is $O(N_{ep} \cdot N_{cand})$ according to Corollary 2. □

　　　For the parameters described in this section, when only the error patterns for the average number $\bar{n}_e \simeq 17.5$ of errors are considered, the computational complexity of deriving the common key is $O(2^{131} \cdot N_{cand})$ according to Lemma 5.

　　　This estimation considers only the error patterns for the average number of errors, while the actual number of bit errors is distributed with a standard deviation of $\sigma_2$ around $\bar{n}_e$. We need to consider all possible error patterns, and their probabilities of occurring as well. The number of patterns when each probability is different can be estimated using the Shannon entropy. For example, an entropy of 131 bits effectively corresponds to $2^{131}$ error patterns.

　　　The parameters used in coding are controlled such that Bob can correct any errors. The number of bit errors is in the error-correctable region, i.e., $\Pr\{n_e > (d -$ $1)/2\} \ll 1$. In this case, Lemma 6 is satisfied.

**Lemma 6:** Let us assume that only a random number sequence and parity check symbols are given in Method 1 using Coding 1. Let $n_e$ be the number of bit errors in one block. If $\Pr\{n_e > (d-1)/2\} \ll 1$ is satisfied, the computational complexity of deriving at least one bit of the common key is $O(2^{H_p} N_{cand})$, where $H_p \simeq mk \cdot h(p_E)$.

**Proof:** The number of error patterns in one block is $\binom{mk}{n_e}$ for $n_e$ bit errors, and the occurrence probability of each error pattern is $p_n = p_E^{n_e}(1-p_E)^{mk-n_e}$. The Shannon entropy of the error-correctable region, i.e., $0 \le n_e \le (d-1)/2$, is $H_p = -\sum_{n_e=0}^{\lfloor (d-1)/2 \rfloor} \binom{mk}{n_e} p_n \log_2 p_n$. If $\Pr\{n_e > (d-1)/2\} \ll 1$ is satisfied, $-\sum_{n_e=0}^{\lfloor (d-1)/2 \rfloor} \binom{mk}{n_e} p_n \log_2 p_n \gg -\sum_{n_e=\lfloor (d-1)/2 \rfloor+1}^{mk} \binom{mk}{n_e} p_n \log_2 p_n$; therefore, we obtain $H_p \simeq -\sum_{n_e=0}^{mk} \binom{mk}{n_e} p_n \log_2 p_n$ by extending the region of the sum to $mk$. This quantity considers all error patterns for an $mk$ bit sequence. In this case, it is equal to the equivocation for $mk$ bits, and $H_p \simeq mk \cdot h(p_E)$. Thus, the computational complexity of deriving at least one bit of the common key is $O(2^{H_p} N_{cand})$, and $H_p \simeq mk \cdot h(p_E)$. □

This computational complexity can be checked by making the following rough estimate. Suppose $m = 8$, $k = 167$, and $p_{eff} = 0.1$ ($p_E \simeq 0.0131$); then $H_p \simeq 134$. From $\sigma_2 \simeq 4.15$ and $\binom{mk}{\bar{n}_e} \simeq 1.0 \times 2^{131}$, we find that $2\sigma_2 \binom{mk}{\bar{n}_e} \simeq 2^{H_p}$, and $2^{H_p}$ is surely the effective number of error patterns.

The truth or falseness of each candidate can be judged by decoding a sufficient number of blocks with the candidate common key as follows: When a candidate is true, the number of bit errors is distributed around $\bar{n}_e$ in all blocks, and parity check symbols are never an error. In contrast, parity check symbols can be an error when a candidate is false. Moreover, in this case, because the information symbols become a haphazard sequence, the code word vector for it is probabilistically uniformly spread out over the code word vector space, and the number of bit errors is uniformly distributed throughout the correctable error numbers. Thus, each candidate can be judged as being true or false from the distribution of errors if a sufficient number of blocks are checked.

**Lemma 7:** Let us assume that only random number sequence and parity check symbols are given in Method 1 using Coding 1. The computational complexity of deriving the information symbols of one block, i.e., $x_1$ ($x_1$*), is equal to that of deriving at least one bit of the common key.

**Proof:** As shown in the proof of Lemma 4 and the following paragraph, the process of deriving the common key $k_e$ consists of listing the candidates of information symbols and $k_e$, checking each candidate $k_e$ using other blocks, and obtaining the final solution. The process of deriving the information symbols of a target block also consists of listing the candidates of information symbols and checking them. To check them, the candidates of $k_e$ are listed and each candidate $k_e$ is checked using other blocks. It is when the final solution of $k_e$ is confirmed that the candidate of the information symbols is confirmed. According to Corollary 1, it is impossible to derive only part of the common key. Thus, Lemma 7 is satisfied. □

6.2 Deriving the common key by using $s_{kA}$*

According to the assumption, Eve obtains not only $z$ ($z$*), $c_I$ ($c_I$*), and $c_{II}$ ($c_{II}$*) but also $s_{kA}$* that is not used message transmissions. Can $s_{kA}$* ease deriving the common key $k_e$? The following Lemma 8 sweeps away this concern.

The conditional secrecy capacity is $C_s \ge C_{s0} = (k - t_m)/n \cdot h(p_{\sigma E}) - n_s/un$ for binary coding from Lemma 1. When the coding is over GF($2^m$), $n$ and $k$ are translated into $mn$ and $mk$, and $t_m = mt_{mc}$. Therefore, $C_s \ge C_{s0} = (k - t_{mc})/n \cdot h(p_{\sigma E}) - n_s/umn$ when the coding is over GF($2^m$). The condition for $n_r$ in Method 1 is translated into $n_r/unm \le C_{s0}$. Thus, $n_r/u \le nmC_{s0} = m(k - t_{mc}) \cdot h(p_{\sigma E}) - n_s/u$. Let $H_s' = n_r/u$ and $H_s = nmC_{s0}$; then, $H_s' \le H_s$. $H_s'$ is the number of secret keys generated per block. In Method 1, $u \ge 1$.

Secret keys are generated from noise. There is a rare case where the number of bit errors is extremely small owing to the statistical fluctuations of noise. We assume that the rare case is bounded with a small quantity $\gamma$, i.e., $\Pr\{n_{ue} < \bar{n}_{ue} - r\sigma_{u2}\} < \gamma$. In addition, we assume that Eve's residual information after the secret key generation is also bounded by the small quantity $\gamma$, i.e., $I(S_{kA}^{n_r}|k_e \leftarrow K_e; S_{kE}^{n_r})/n_r < \gamma$.

**Lemma 8:** In Method 1 using Coding 1, the computational complexity of Eve's deriving at least one bit of the common key is equal to that of deriving it only from a random number sequence and parity check symbols if the effects of $\Pr\{n_{ue} < \bar{n}_{ue} - r\sigma_{u2}\} < \gamma$ and $I(S_{kA}^{n_r}|k_e \leftarrow K_e; S_{kE}^{n_r})/n_r < \gamma$ are negligible, where $\gamma$ is a small quantity.

**Proof:** According to the assumption, Eve can obtain $s_{kA}*$ that is not used for message transmissions. Let us assume that $s_{kA}*$ is generated from the block Eve wants to analyze. If the inverse operation of universal hashing used in generating secret keys were easy for Eve, the information symbols $x_1*$ in that block could be derived, and the number $N_s$ of candidates would satisfy $N_s \geq 2^{mk-H_{s'}}$, where equality corresponds to the case of $u = 1$. When the information symbols are derived by using one block of parity check symbols, the number of candidates is $N_p = 2^{mk-m(n-k)}$ according to Lemma 3. When the error-correcting code works correctly, the amount of redundant information $m(n - k)$ exceeds the entropy of the bit errors $mk \cdot h(p_E)$, i.e., $m(n - k) \geq mk \cdot h(p_E)$. Because of $H_s = m(k - t_{mc}) \cdot h(p_{\sigma E}) - n_s/u$, $mk \cdot h(p_E) > H_s$ is satisfied. Because $H_s \geq H_{s'}$, $m(n - k) > H_{s'}$. Thus, $N_p < N_s$. Next, let us assume that Eve tries to correlate the information in the random-number transmission stage with $s_{kA}*$. However, because $s_{kA}*$ is generated in the capacity of $n_r/u \leq nmC_{s0}$, as long as Eve fails to derive $x_1*$, the information in the random-number transmission stage is uncorrelated with $s_{kA}*$ if the effects of $\Pr\{n_{ue} < \bar{n}_{ue} - r\sigma_{u2}\} < \gamma$ and $I(S_{kA}{}^{n_r}|k_e \leftarrow K_e; S_{kE}{}^{n_r})/n_r < \gamma$ are negligible. (See Fig. 3 and proof (2) of Lemma 1.) Without any correlation, it is advantageous for Eve to use the information in the random-number transmission stage when trying to derive the common key, but not to use $s_{kA}*$ because of $N_p < N_s$. Therefore, Eve will use the information in the random-number transmission stage until she succeeds in deriving $x_1*$. Thus, Lemma 8 is satisfied. □

$\Pr\{n_{ue} < \bar{n}_{ue} - r\sigma_{u2}\}$ can be made exponentially small, as follows. The number of bit errors obeys a binomial distribution. When it is approximated with a normal distribution, $\Pr\{n_{ue} < \bar{n}_{ue} - r\sigma_{u2}\} \simeq (1/\sqrt{2})\int_{-\infty}^{-r} e^{-t^2/2}dt = O(e^{-r^2/2}/r)$, where $t = (n_{ue} - \bar{n}_{ue})/\sigma_{u2}$. Thus, $\Pr\{n_{ue} < \bar{n}_{ue} - r\sigma_{u2}\}$ is exponentially small if $r$ is appropriately chosen. $I(S_{kA}{}^{n_r}|k_e \leftarrow K_e; S_{kE}{}^{n_r})/n_r$ can be also exponentially small. As described in the proof of Lemma 1, $I(S_{kA}{}^{n_r}|k_e \leftarrow K_e; S_{kE}{}^{n_r})/n_r \leq 2^{-n_s}/n_r \ln 2$. Because $n_s$ and $n_r$ can be chosen to be $O(umn)$, when $umn$ is sufficiently large, $2^{-n_s}/n_r \ln 2$ is exponentially small. Thanks to these characteristics, we can choose a sufficiently small $\gamma$.

**Corollary 3:** In Method 1 using Coding 1, the computational complexity of Eve's deriving $x_1*$ is equal to that of deriving it only from a random number sequence and parity check symbols if the effects of $\Pr\{n_{ue} < \bar{n}_{ue} - r\sigma_{u2}\} < \gamma$ and $I(S_{kA}{}^{n_r}|k_e \leftarrow K_e; S_{kE}{}^{n_r})/n_r < \gamma$ are negligible, where $\gamma$ is a small quantity.

**Proof:** The corollary is apparent from the proof of Lemma 8. □

Lemma 9 follows from Lemmas 6 and 8.

**Lemma 9:** In Method 1 using Coding 1, the computational complexity of Eve's deriving at least one bit of the common key $k_e$ is $O(2^{H_P} N_{cand})$ if the effects of $\Pr\{n_{ue} < \bar{n}_{ue} - r\sigma_{u2}\} < \gamma$ and $I(S_{kA}{}^{n_r}|k_e \leftarrow K_e; S_{kE}{}^{n_r})/n_r < \gamma$ are negligible, where $\gamma$ is a small quantity.

**Corollary 4:** The effective key length in Method 1 using Coding 1 is $N_K - m(n - k) + mk \cdot h(p_E) + \log_2(1 - \delta)$.

**Proof:** The corollary is apparent from $\log_2(2^{H_P} N_{cand}) = N_K - m(n - k) + mk \cdot h(p_E) + \log_2(1 - \delta)$. □

**Corollary 5:** In Method 1 using Coding 1, the computational complexity of Eve's deriving $x_1$ ($x_1*$) is $O(2^{H_P} N_{cand})$ if the effects of $\Pr\{n_{ue} < \bar{n}_{ue} - r\sigma_{u2}\} < \gamma$ and $I(S_{kA}{}^{n_r}|k_e \leftarrow K_e; S_{kE}{}^{n_r})/n_r < \gamma$ are negligible, where $\gamma$ is a small quantity.

**Proof:** The corollary is apparent from Corollary 3 and Lemmas 7 and 9. □

**Lemma 10:** In Method 1 using Coding 1, the computational complexity of Eve's deriving at least one bit of the secret key $s_{kA}$ is at least $O(2^{H_P} N_{cand})$ if the effects of $\Pr\{n_{ue} < \bar{n}_{ue} - r\sigma_{u2}\} < \gamma$ and $I(S_{kA}{}^{n_r}|k_e \leftarrow K_e; S_{kE}{}^{n_r})/n_r < \gamma$ are negligible, where $\gamma$ is a small quantity.

**Proof:** Because secret keys are generated from $x_1$ in units of $u$ blocks, when Eve derives at least one bit of the secret key $s_{kA}$, she needs $x_1$ for $u$ blocks; moreover, she needs to perform algorithm $S$ for generating the secret keys. From Corollary 5, the computational complexity of only deriving one block of $x_1$ is $O(2^{H_P} N_{cand})$ if the effects of $\Pr\{n_{ue} < \bar{n}_{ue} - r\sigma_{u2}\} < \gamma$ and $I(S_{kA}{}^{n_r}|k_e \leftarrow K_e; S_{kE}{}^{n_r})/n_r < \gamma$ are negligible. To derive at least one bit of $s_{kA}$, algorithm $S$ must be analyzed moreover. Thus, Lemma 10 is satisfied. □

**Lemma 11:** In Method 1 using Coding 1, encoding $E_a$ is computationally secure in the sense of Definition 3.

**Proof:** Let $\eta = 1/\gamma$. From Corollary 5, the computational complexity of Eve's deriving $x_1$ ($x_1*$) is $\{O(2^{H_P} N_{cand})[1 - O(1/\eta)] + O(1/\eta)\}$ by taking into account $\Pr\{n_{ue} < \bar{n}_{ue} - r\sigma_{u2}\} < \gamma$ and $I(S_{kA}{}^{n_r}|k_e \leftarrow K_e; S_{kE}{}^{n_r})/n_r < \gamma$, where $\gamma$ is a small quantity; the term $O(1/\eta)$ comes from those rare cases, and the term $O(2^{H_p} N_{cand})[1 - O(1/\eta)]$ comes from the other cases. Thus, the probability of successfully guessing information symbols, $p_d$ in Definition 3, is $p_d \leq 1/2^k + 1/\{O(2^{H_p} N_{cand})[1-O(1/\eta)]+O(1/\eta)\}$. The parameter $mk$ is determined such that it satisfies $\overline{N_0} + \lfloor r_1\sigma_1 \rfloor \leq mk$, and thus, $mk = O(N_K)$. Moreover, $mn = O(mk)$. Thus, $N_K - m(n-k) = O(N_K)$. As is apparent from $N_{cand} = 2^{N_K-m(n-k)}(1-\delta)$, the parameters $N_K$, $m$, $n$ and $k$ are chosen such that $N_K - m(n-k) > 0$. In summary, $N_K - m(n-k) = O(N_K) > 0$. Hence, $N_{cand} = 2^{N_K-m(n-k)}(1-\delta) > P(N_K)$ is satisfied at $N_K \to \infty$ for every polynomial equation $P(N_K)$. In addition, $\gamma \to 0$ can be chosen for $N_K \to \infty$. Therefore, when $k_0$ is chosen sufficiently large, $p_d < 1/2^k + 1/P(N_K)$ is satisfied for $N_K \geq k_0$. Thus, encoding $E_a$ in Method 1 using Coding 1 is computationally secure in the sense of Definition 3. □

The following theorem is obtained from Lemma 10.

**Theorem 1:** Method 1 using Coding 1 is computationally secure in the sense of Definition 2.

**Proof:** Let $\eta = 1/\gamma$. From Lemma 10, the computational complexity of Eve's deriving any one bit of the secret key $s_{kA}$ is at least $\{O(2^{H_p} N_{cand})[1 - O(1/\eta)] + O(1/\eta)\}$ by taking into account $\Pr\{n_{ue} < \overline{n}_{ue} - r\sigma_{u2}\} < \gamma$ and $I(S_{kA}^{n_r}|\mathbf{k}_e \leftarrow \mathbf{K}_e; S_{kE}^{n_r})/n_r < \gamma$, where $\gamma$ is a small quantity. Therefore, the probability of successfully guessing the secret key, $p_s$ in Definition 2, is $p_s \leq 1/2 + 1/\{O(2^{H_p} N_{cand})[1-O(1/\eta)]+O(1/\eta)\}$. The parameter $mk$ is determined such that it satisfies $\overline{N_0} + \lfloor r_1\sigma_1 \rfloor \leq mk$, and thus, $mk = O(N_K)$. Moreover, $mn = O(mk)$. Thus, $N_K - m(n-k) = O(N_K)$. As is apparent from $N_{cand} = 2^{N_K-m(n-k)}(1-\delta)$, the parameters $N_K$, $m$, $n$ and $k$ are chosen such that $N_K - m(n-k) > 0$. In summary, $N_K - m(n-k) = O(N_K) > 0$. Hence, $N_{cand} = 2^{N_K-m(n-k)}(1-\delta) > P(N_K)$ is satisfied at $N_K \to \infty$ for every polynomial equation $P(N_K)$. In addition, $\gamma \to 0$ can be chosen for $N_K \to \infty$. Therefore, when $k_0$ is chosen sufficiently large, $p_s < 1/2 + 1/P(N_K)$ is satisfied for $N_K \geq k_0$. Thus, Method 1 using Coding 1 is computationally secure in the sense of Definition 2. □

The redundant information $x_2$ is transmitted through a public channel in Method 1. This is to make the security analysis easy. However, an actual system might transmit $x_2$ through the same channel as that for $x$. For this reason, the following Method 2 is defined.

**Method 2:** In this modification of Method 1, $x_2$ is transmitted through the same channel as $x$ (See Fig. 6).

In this case, bit errors occur in $x_2$, and deriving the common key is more difficult than that in Method 1. Therefore, the claim of Theorem 1 is true for Method 2.

**Corollary 6:** Method 2 using Coding 1 is computationally secure in the sense of Definition 2.

Lemma 11, Theorem 1, and Corollary 6 can be proved without assuming any mathematical difficulties. This means that Methods 1 and 2 using Coding 1 face no threat that an efficient decrypting algorithm might be found.

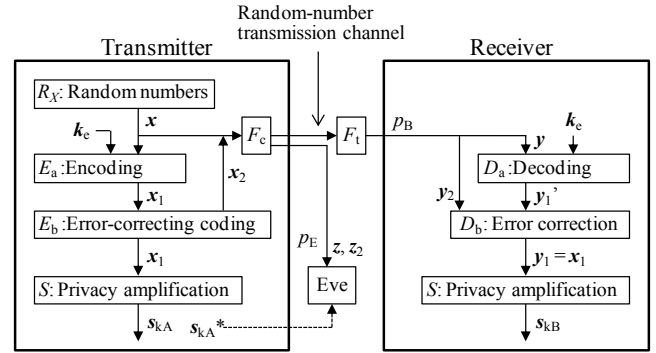

**Fig. 6** Block diagram of Method 2.

## 7. Design example

### 7.1 Parameters

Table 1 summarizes Method 1 using Coding 1. As mentioned in section 5.3, the parameters should be determined such that $\overline{N_0} + \lfloor r_1\sigma_1 \rfloor \leq mk$ in order to prevent Eve from deriving the common key part-by-part. For example, given a (255, 167) linear code over $GF(2^8)$ and $N_K = 2496$, then $\overline{N_0} = 1248$, $mk = 1336$, and $\sigma_1 \simeq 24.98$; in this case, $\overline{N_0} + \lfloor r_1\sigma_1 \rfloor = 1335 < mk$ is satisfied for $r_1 = 3.5$. In other words, we can choose $N_K = 2496$ in this code. The computational complexity of deriving the common key is proportional to $2^{H_p} N_{cand} \simeq 2^{mk \cdot h(p_E)} 2^{N_K - m(n-k)}(1-\delta) \simeq 2^{1926}(1-\delta)$ for $p_E \simeq 0.0131$ due to Lemma 6. This value seems to be sufficiently large; if a larger value is required, we can enlarge the block size.

So far, we have used the common key repeatedly. This is because we wanted to evaluate the basic performance of the proposed method. However, if we use the common key more cryptographically, i.e., as a seed key of pseudo-random numbers, the security will increase computationally. Moreover, bit errors will still enhance security. In the analysis described in section 6.1.2, the security was determined by only the complexity of the analysis for one block. This is because the common key is repeatedly used. However, if pseudo-random numbers are used instead, the number of blocks needed to derive the common key increases. Let the needed number of blocks be $N_T$. In this case, the entropy of noise that affects the analysis is $N_T \cdot mk \cdot h(p_E)$, and the effective key length increases. This effect is powerful because it is information theoretic. For the above example, where $m(n - k) = 704$ and $H_p \simeq 134$, $2^{N_T H_p} N_{cand} > 2^{N_K}(1-\delta)$ is satisfied for $N_T \geq 6$. In this case, an exhaustive search of $\mathcal{K}_e$ is needed for deriving the common key. Moreover, the following observations can be made. We assumed that parity check symbols exactly leak to Eve in Method 1. However, when we use Method 2, it adds bit errors to the parity check symbols, and consequently, its security increases. Moreover, $x$ and $x_2$ are transmitted without encryption in Methods 1 and 2. If $x$ and $x_2$ are encrypted with pseudo-random numbers, security increases computationally, although another key is needed.

**Table 1** Summary of Method 1 using Coding 1

|  | Example |
|---|---|
| (0) Share common key | $N_K = 2496$ |
| $\mathcal{K}_e = \{k_e \in \{0, 1\}^{N_K} \mid -r_1\sigma_1 \leq N_1(k_e) - \overline{N}_0 \leq r_1\sigma_1\}$ | $\overline{N}_0 = 1248$ |
| $N_1 + N_2 = N_K$ | $r_1 = 3.5$ |
|  | $\sigma_1 \simeq 24.98$ |
| (1) Generate random number sequence |  |
| $x \leftarrow R_X$ |  |
| (2a) Divide random number sequence | Fig. 1 |
| $x^{(N_K)} \times k_e \mapsto b_I^{(N_1)} \times b_{II}^{(N_2)}$ |  |
| (2b) Perform $(n, k)$ block coding: | RS code over GF($2^m$) |
| $b_I \mapsto c_I$ and $b_{II} \mapsto c_{II}$ | $n = 255$ |
| $\overline{N}_0 + \lfloor r_1\sigma_1 \rfloor \leq mk$ | $k = 167$ |
|  | $m = 8$ |
| (3) Generate secret key (Privacy amplification) | Table II |

### 7.2 Conditional secrecy capacity

Let us estimate an example of the conditional secrecy capacity when using Method 1, a (255, 167) linear MDS code over GF($2^8$), and $P_{eff} = 0.1$ ($p_E \simeq 0.0131$).

As described in section 6.2, when the coding is over GF($2^m$), $C_s \geq (k - t_{mc})/n \cdot h(p_{\sigma E}) - n_s/umn$. If $u = 1$ and $r = 3$, then $\overline{n}_{ue} = umkp_E \simeq 17.5$, $\sigma_{u2} = \sqrt{umkp_E(1 - p_E)} \simeq 4.15$, $p_{\sigma E} = p_E \cdot (\overline{n}_{ue} - r\sigma_{u2})/\overline{n}_{ue} \simeq 0.00376$, and $h(p_{\sigma E}) \simeq 0.0357$. Thus, $C_s \geq (k - t_{mc})/n \cdot h(p_{\sigma E}) - n_s/umn \simeq 0.0111 - n_s/umn$. If we choose $n_s = 10$, then $C_s \geq 0.00615$ and $n_r \geq 12.5$. Here, "$\geq$" is used to indicate a lower bound.

Let us determine $\gamma$ by referring to the above values, although this process is the inverse of that from the viewpoint of the meaning that $\gamma$ should be given first. The condition in Definition 4 (1) is $\Pr\{S_{kA}^{n_r} \neq S_{kA}^{n_r} \mid k_e \leftarrow K_e\} < \gamma$. Let $n_{cb}$ be the number of Bob's code errors in one block. When $u = 1$, then $\Pr\{S_{kA}^{n_r} \neq S_{kB}^{n_r} \mid k_e \leftarrow K_e\} = \Pr\{n_{cb} > (n - k)/2\}$. The method in this report works efficiently when $p_B - p_E \ll p_E$ (see section 7.5).

Therefore, let us assume $p_B = p_E$ as an example. In this case, $\Pr\{n_{cb} > (n - k)/2\} < 4.70 \times 10^{-10}$, where the third decimal place is rounded up. Definition 4 (2) requires $I(S_{kA}^{n_r}\mid k_e \leftarrow K_e; S_{kE}^{n_r})/n_r < \gamma$, except for the rare case of $n_{ue}/umk < p_{\sigma E}$. The probability of the rare case is $\Pr\{n_{ue}/umk < p_{\sigma E}\} < 4.48 \times 10^{-4}$ for $P_{eff} = 0.1$, and $I(S_{kA}^{n_r}\mid k_e \leftarrow K_e; S_{kE}^{n_r})/n_r \leq 2^{-n_s}/n_r \ln 2 < 1.13 \times 10^{-4}$. From the above three kinds of small values, $\gamma \leq \max(4.70 \times 10^{-10}, 4.48 \times 10^{-4}, 1.13 \times 10^{-4}) = 4.48 \times 10^{-4}$. Table I summarizes these values.

$C_s$ increases as $u$ increases, and $\gamma$ can be decreased as $r$ and $n_s$ are increased. If $u = 10$ and $r = 5$, then $\overline{n}_{ue} = umkp_E \simeq 175$, $\sigma_{u2} = \sqrt{umkp_E(1 - p_E)} \simeq 13.1$, $p_{\sigma E} = p_E \cdot (\overline{n}_{ue} - r\sigma_{u2})/\overline{n}_{ue} \simeq 0.00817$, and $h(p_{\sigma E}) \simeq 0.0684$. Thus, $C_s \geq 0.0212 - n_s/umn$. Here, if $n_s = 16$, then $C_s \geq 0.0204$, $n_r \geq 416$. Table I lists $\gamma$–related values. It also shows the case of $u = 10$, $r = 3$, and $n_s = 10$. When $u \to \infty$, then $h(p_E) \simeq 0.101$ and $C_s \geq 0.0312$.

**Table 2** Lower bound of conditional secrecy capacity and related quantities at $(n, k) = (255, 167)$, $m = 8$, and $P_{eff} = 0.1$ ($p_E \simeq 0.0131$). (1) is related to Definition 4 (1); (2.1) is related to Definition 4 (2) and (4); (2.2) is related to Definition 4 (2). The third decimal place is rounded up in those rows. $n_{cb}$ denotes the number of Bob's code errors in one block. $n_{ue}$ denotes the number of Eve's bit errors in $u$ blocks.

| | | | |
|---|---|---|---|
| $u$ | 1 | 10 | 10 |
| $r$ | 3 | 3 | 5 |
| $n_s$ | 10 | 10 | 16 |
| (1) $\Pr\{n_{cb} > (n - k)/2\} <$ | $4.70 \times 10^{-10}$ | $4.70 \times 10^{-9}$ | $4.70 \times 10^{-9}$ |
| (2.1) $\Pr\{n_{ue}/mk < p_{\sigma E}\} <$ | $4.48 \times 10^{-4}$ | $9.63 \times 10^{-4}$ | $5.07 \times 10^{-8}$ |
| (2.2) $2^{-n_s}/n_r \ln 2 <$ | $1.13 \times 10^{-4}$ | $2.79 \times 10^{-6}$ | $5.29 \times 10^{-8}$ |
| $\gamma \leq$ | $4.48 \times 10^{-4}$ | $9.63 \times 10^{-4}$ | $5.29 \times 10^{-8}$ |
| $C_s \geq$ | 0.00615 | 0.0248 | 0.0204 |
| $n_r/u \geq$ | 12.5 | 50.6 | 41.6 |

### 7.3 Multiple codes

Two kinds of error-correcting codes are often combined to make the error correction perfect, e.g., product codes and concatenated codes [23]. The method in this report can be modified to suit double coding using two kinds of common keys. A concrete example is as follows.

The random numbers, amounting to $N_b$ blocks of groups I and II, of a sequence coded using $k_e$ are shuffled, and the shuffled sequence is then coded using another common key $k_{ed}$. The parameters for the two codes do not need to be the same. Decoding is possible from either the $k_e$- or $k_{ed}$-related code, and this double coding is resistant to burst errors. For example, let us decode the $k_e$-related code first and assume there are residual errors. Because the random number sequence is shuffled, the residual errors are distributed over multiple blocks in the $k_{ed}$-related code. They can be corrected through $k_{ed}$-related error correction. Here, although the shuffling process becomes computationally expensive, $N_b$ should be as large as possible. The value of $N_b$ should be determined on the basis of the processing performance of the transmitter and receiver.

The double coding is for complete error correction, but there is a possibility that all errors will be corrected in one decoding. The security of this method, therefore, is quantified by the complexity of the decryption process of one of the two codes. An important thing in double coding is to prevent the parity check symbols in one of the two codes from affecting the complexity of the decryption process in the other code.

As mentioned in section 6.1.1, the process of deriving the common key includes listing the candidates of the information symbols for one block as a basic component. Because this listing is a closed process for one block, the parity check symbols in the $k_e$ ($k_{ed}$)-related code do not contribute to the process of listing the candidates of the information symbols in the $k_{ed}$ ($k_e$)-related code. Therefore, the security of this method is determined by the complexity of deriving only one of the common keys. However, the conditional secrecy capacity changes. Here, let the block size be the same for both codings. Because redundant information on the $k_e$ ($k_{ed}$)-related code can correct $t_m$ ($t_{md}$) bits at maximum, the conditional secrecy capacity is $C_s \geq (k - t_m - t_{md})/n \cdot h(p_{\sigma E}) - n_s/un$. Here, $t_m$ and $t_{md}$ can be set less than those of single coding thanks to double coding.

### 7.4 Noise source

The output of an LD used in optical communications includes noise; the phase of the output light is especially noisy and is sufficiently random [15]. Coding methods like Phase-Shift Keying (PSK) or Differential Phase-Shift Keying (DPSK) use the phase of light. Thus, the method in this report can use phase noise-related bit errors as a resource that is always available in optical communications.

### 7.5 BER in random-number transmission channel

As mentioned in the preceding subsection, the phase noise of an LD output is directly usable in optical communications. However, when the environmental noise $F_t$ in a transmission channel is large, the condition $p_E \simeq p_B$ ($p_B - p_E \ll p_E$) is not satisfied, where much redundant information is required and the conditional secrecy capacity decreases. One solution in this case is to code the transmitter output $F_c(X)$ with an error-correcting code and to build a pseudo-errorless channel ($p_E = p_B$). Because the purpose of this coding is to transmit a random number sequence with errors correctly, the decoded sequence has errors, and Eve does not obtain any new information.

### 8. Summary

Secure communications using noise generally need a mechanism to make Eve less advantageous than Bob. However, such a mechanism does not always exist intrinsically. This report described an extrinsic method that makes Eve disadvantageous by using a common key. The common key, error-correcting code, and noise are managed in a cooperative manner, and the secret keys are generated from noise. Messages are encrypted with the secret keys by using a one-time pad. As a result, information leaks that are meaningful to Eve are restricted to the parity-check symbols for the random numbers. It is possible to derive the candidates of the common key from the parity check symbols, and the security of this method can be quantified in terms of the computations needed for an exhaustive search of the candidates. We calculated the number of the candidates of the common key by assuming all parity check symbols were leaked to Eve without bit errors. The number is $2^{H_p} N_{cand}$, and it determines the security of this method. Its logarithm $N_K - m(n - k) + mk \cdot h(p_E) + \log_2(1 - \delta)$ corresponds to the effective key length. Methods with computational security generally face the threat that an efficient decryption method might be found. However, this method does not rely on any mathematical difficulties, and therefore, there is no threat that a more efficient decryption method than an exhaustive search might be found. The method requires listing the information symbols from the parity check symbols followed by listing the candidates of the common key in decryption. This threat-less form of security can be used to protect highly confidential information like government and military secrets, although its security level is computational. However, it requires privacy amplification to assure high security, and this reduces the message transmission rate to $R_m \ll 1$.

### Acknowledgments

The author thanks Hisayoshi Sato, Keisuke Hakuta, Tomohiko Uyematsu, and Masashi Ban for their insightful comments.